\documentclass[reprint,superscriptaddress]{revtex4-1}

\usepackage{amsmath}
\usepackage{amsfonts}
\usepackage{amssymb}
\usepackage{graphicx}
\usepackage{moreverb}
\usepackage{url}
\usepackage{epstopdf}
\usepackage{color}
\usepackage{multirow}
\usepackage{subfigure}
\usepackage{textgreek}
\usepackage[title,titletoc,toc]{appendix}

\begin{document}

\title{Continuous melting through a hexatic phase in confined bilayer water}
\author{Jon~Zubeltzu}
\email[E-mail address: ]{jzubeltzu@nanogune.eu}
\affiliation{CIC nanoGUNE, 20018 Donostia-San Sebasti\'{a}n, Spain}
\author{Fabiano~Corsetti}
\affiliation{CIC nanoGUNE, 20018 Donostia-San Sebasti\'{a}n, Spain}
\affiliation{Department of Materials and the Thomas Young Centre for Theory and Simulation of Materials, Imperial College London, London SW7 2AZ, United Kingdom}
\author{M.~V.~Fern\'andez-Serra}
\affiliation{Physics and Astronomy Department, SUNY Stony Brook University, New York 11794-3800, USA}
\author{Emilio~Artacho}
\affiliation{CIC nanoGUNE, 20018 Donostia-San Sebasti\'{a}n, Spain}
\affiliation{Theory of Condensed Matter, Cavendish Laboratory, University of Cambridge, Cambridge CB3 0HE, United Kingdom}
\affiliation{Basque Foundation for Science Ikerbasque, 48011 Bilbao, Spain}
\affiliation{Donostia International Physics Center, 20018 Donostia-San Sebasti\'{a}n, Spain}
\date{\today}

\begin{abstract}
Liquid water is not only of obvious importance but also extremely intriguing, displaying many anomalies that still challenge our understanding of such an {\em a priori} simple system. The same is true when looking at nanoconfined water: The liquid between constituents in a cell is confined to such dimensions, and there is already evidence that such water can behave very differently from its bulk counterpart. A striking finding has been reported from computer simulations for two-dimensionally confined water: The liquid displays continuous or discontinuous melting depending on its density. In order to understand this behavior, we have analyzed the melting exhibited by a bilayer of nanoconfined water by means of molecular dynamics simulations. At high density we observe the continuous melting to be related to the phase change of the oxygens only, with the hydrogens remaining liquid-like throughout. Moreover, we find an intermediate hexatic phase for the oxygens between the liquid and a triangular solid ice phase, following the Kosterlitz-Thouless-Halperin-Nelson-Young theory for two-dimensional melting. The liquid itself tends to maintain the local structure of the triangular ice, with its two layers being strongly correlated, yet with very slow exchange of matter. The decoupling in the behavior of the oxygens and hydrogens gives rise to a regime in which the complexity of water seems to disappear, resulting in what resembles a simple monoatomic liquid. This intrinsic tendency of our simulated water may be useful for understanding novel behaviors in other confined and interfacial water systems.
\end{abstract}

\maketitle
\section{Introduction}

The dimensionality of a material is a crucial factor that strongly determines its properties. Water is no exception: Its behavior changes significantly from the bulk when it is under one-dimensional (1D) or 2D nanoconfinement. Nanoconfined water has been investigated in detail in the past few decades by many experimental (1D~\cite{reiter2012, mallamace2007, wang2008}, 2D~\cite{algara2015, ortiz2013, reiter2013}) and theoretical (1D~\cite{Xu2011,javadian2012,klameth2013}, 2D~\cite{zangi2004,corsetti2015a,zangi2003,kaneko2014,vilanova2011,de2012,strekalova2011,de2011,mazza2012,chen2015,mario2015,han2010,corsetti2015b,bai2012,giovambattista2009,bai2003,slovak1999,krott2013,johnston2010,koga1997,giovambattista2006,kumar2005,meyer1999,kumar2007,mosaddeghi2012}) studies for two main reasons: First, it appears in many biological, geological, and nanotechnological systems~\cite{bertrand2013}, playing a very active role. Second, its study will help complete the understanding of the complex phenomenology that is known for bulk water~\cite{ball2008}.

Algara {\em et al.}~\cite{algara2015} have recently observed experimentally that water is structured into square ice made of different numbers of layers depending on the available space between two graphene sheets at ambient temperature. Computational studies agree that under two-dimensional nanoconfinement water is structured into layers perpendicularly oriented with respect to the confining direction (monolayer \cite{zangi2004,corsetti2015a,zangi2003,kaneko2014,vilanova2011,de2012,strekalova2011,de2011,mazza2012}, bilayer~\cite{chen2015,mario2015,han2010,corsetti2015b,bai2012,giovambattista2009,bai2003,slovak1999,krott2013,johnston2010,koga1997}, trilayer and so on~\cite{giovambattista2006,kumar2005,meyer1999,kumar2007,mosaddeghi2012}). Most of the computational studies agree with the existence of a stable monolayer square ice phase~\cite{zangi2004,corsetti2015a,zangi2003,kaneko2014}. For the bilayer and trilayer cases, although there are recent studies getting the square ice~\cite{chen2015,mario2015}, still the majority obtain different types of structure as the most stable ones. Han {\em et al.}~\cite{han2010} observed by classical molecular dynamics simulations using the TIP5P force-field model~\cite{mahoney2000} that nanoconfined bilayer water can freeze by two types of phase transitions depending on its density: a first-order phase transition into honeycomb ice at low densities and a continuous phase transition into rhombic ice at high densities. A recent work by Corsetti {\em et al.}~\cite{corsetti2015b} based on molecular dynamics using the TIP4P/2005 force-field model~\cite{abascal2005} and density-functional theory, however, distinguishes two different stable bilayer ices at high densities: a proton-ordered rhombic phase for low temperatures and a proton-disordered triangular phase for high temperatures. 

In this work, we analyze the behavior of bilayer liquid water and its different melting phase transitions by computer simulations. We observe that at low densities water freezes into honeycomb ice as previously reported. At high densities, however, depending on temperature, water can freeze via first-order phase transition into a proton-ordered rhombic ice or via continuous phase transition into a proton-disordered triangular ice. The observation of an intermediate hexatic phase between the higher-temperature ice and the liquid phases suggest that the continuous phase transition fits the Kosterlitz-Thouless-Halperin-Nelson-Young (KTHNY) melting theory. The liquid near the continuous phase transition maintains the same features of the triangular ice: strong layering, interlayer correlation, local triangular structure, and a decoupling between the dynamics of oxygens and hydrogens. These features give rise to an unexpected conclusion: There is an area in the phase diagram of bilayer water in which the effective behavior can be mapped onto that of a simple monoatomic fluid.

\section{Methods}
We carry out computational simulations based on classical molecular dynamics (MD) and {\em ab initio} molecular dynamics (AIMD). For the former, we use the LAMMPS code~\cite{plimpton1995} and the TIP4P/2005 force field to model the interaction among the water molecules, where the cutoff of the Lennard-Jones interaction is set to 12~\AA\ and the particle-particle particle-mesh (PPPM) method~\cite{hockney1988} is used to compute the long-range Coulombic interaction. During the first 60 ns of the MD we adopt the constant particle number, volume, and temperature ensemble ({\em NVT}) and use the Berendsen thermostat in order to control the temperature of the system. Then, the constant particle number, volume, and energy ensemble ({\em NVE}) is used for 2 ns and the data are collected. The size of the square cell with $\left ( 34.90 \times 34.90 \right)$~\AA$^2$ dimensions is fixed, and the number of molecules ranges from 196 to 314 in order to sample different densities. For system size testing, see Ref.~\cite{corsetti2015b}.

For the AIMD calculations based on density functional theory, we use the SIESTA code~\cite{soler2002} with a fully non local exchange and correlation~\cite{dion2004}, devised to describe the van der Waals interactions for water (as in Ref.~\cite{corsetti2015a}). The final configuration obtained from the MD calculations is annealed for 5 ps and then the {\em NVE} ensemble is used for at least 10 ps by AIMD while data are collected. Due to the larger computational cost of such calculations, we reduce the size of the cell to $\left ( 23.46 \times 24.47 \right)$~\AA$^2$, and we sample three different densities with 120, 130, and 140 water molecules.

In general for all the calculations, the timestep is set to 0.5 fs and an ({\em xy}) flat Lennard-Jones 9-3 potential is used to confine water along the {\em z} direction that mimics the interaction of water with solid paraffin~\cite{lee1984}: $\epsilon$ = $1.25$ kJ/mole and $\sigma$ = $0.25$ nm. The distance between the confining walls is set to 8~\AA, ensuring a bilayer structure.

\section{Results and Discussion}
\begin{figure}
\includegraphics[width=0.51\textwidth]{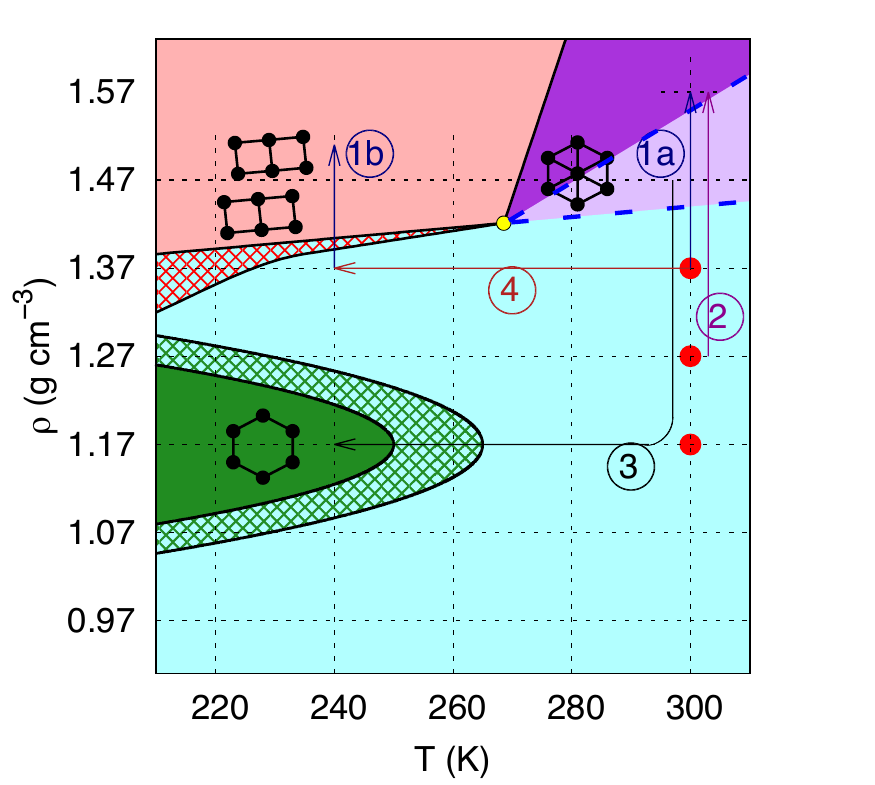}
\caption{Summary of calculations and phase diagram constructed from the results. The density is defined as in Ref.~\cite{han2010}. The crossing points of the thin dashed grid are the points on the phase diagram sampled by MD, while the three red circles show the points also calculated by AIMD. In the areas filled by a crosshatch we observe the coexistence of both liquid and solid phases, consistent with having a first-order phase transition in an {\em NVT} ensemble. The black lines that delimit the low temperature solid phases represent the first-order transition lines, while the blue dashed lines at high temperatures delimit the continuous phase transition lines among the liquid, hexatic and triangular ice solid phases. These transition lines were drawn semi quantitatively from the results obtained at the sampled points. The connection point that joins the first-order and continuous phase transition lines is located at $\left ( 270 \pm 10 \text{ K}, 1.42 \pm 0.05 \text{ g~cm}^{-3}\right)$. The numbered arrows will be useful to understand the results that are shown on the remaining part of the paper.}
\label{fig:pd}
\end{figure}

Figure~\ref{fig:pd} shows a summary of the calculations carried out, together with the phase diagram constructed from the results obtained. At each point on the density-temperature phase diagram we use five different indicators to assign a phase: the oxygen-oxygen radial distribution function (RDF), the diffusion of the oxygens, and the mean positions of the oxygens, the three of them in the {\em xy} plane, the density profile along the confining direction, and the oxygen-oxygen-oxygen angular distribution function. At high densities and temperatures, we have distinguished the hexatic phase (light purple area) from the triangular ice phase (dark purple area), using several indicators which we shall describe in detail later. Our phase diagram is directly comparable with the one obtained by Han {\em et al.}~\cite{han2010} with the TIP5P model. One clear difference is that the present results are shifted towards lower temperatures ($\Delta$T $\sim 40$ K), which is consistent with the fact that TIP5P tends to be more structured than TIP4P/2005~\cite{vega2011}. Around $\rho = 1.17$ g~cm$^{-3}$, corresponding to four atomic layers of (001) Ih ice, we observe the existence of honeycomb ice at low temperatures, in agreement with previous results~\cite{han2010}. One layer of honeycomb ice results from the squashing of two (001) atomic planes of Ih ice into one layer. The most striking change occurs at high densities: Instead of one rhombic phase, two different solid phases are observed. The one at higher temperatures is a proton-disordered triangular ice with close-packed O planes and a very high value of H-configurational entropy (twice that of bulk ice~\cite{corsetti2015b}). The ice at lower temperatures is a proton-ordered rhombic ice, that is characterized by the formation of square-shape tubes with fixed position of both the oxygens and the hydrogens, and no bonding between the tubes. These two solids are connected by an order-disorder first-order phase transition (for a more detailed discussion of these phases we refer to Ref.~\cite{corsetti2015b}). 

\begin{figure}[b!]
\includegraphics[width=0.5\textwidth]{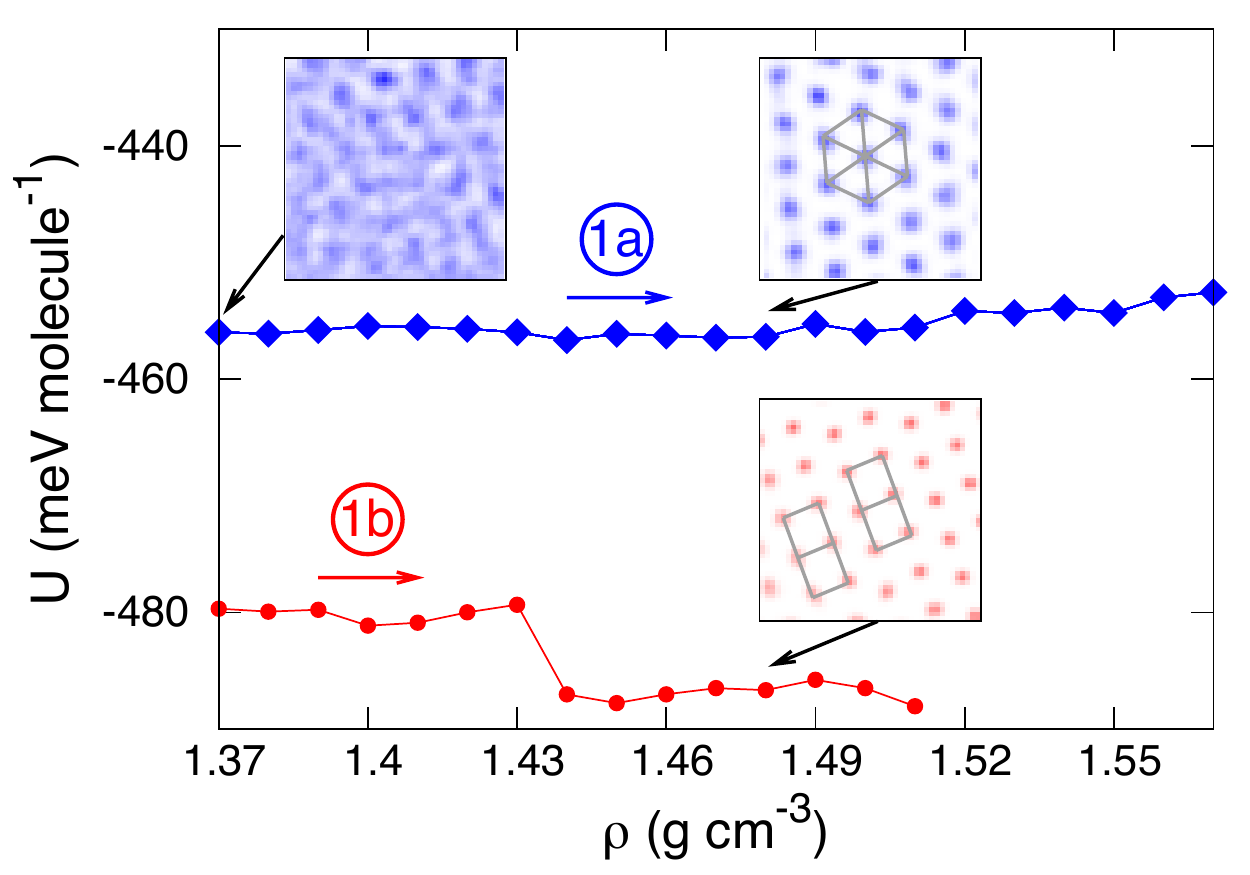}
\caption{Potential energy as a function of density for $T = 240$ K (bottom, red) and $T = 300$ K (top, blue). The 1a and 1b arrows refer to the paths on the phase diagram (see Fig.~\ref{fig:pd}). The insets show the averaged positions of the oxygens during 100 ps in a window of dimensions $\left ( 15 \times 15 \right)$~\AA$^2$ within the cell. The gray lines within the high-density insets are drawn to illustrate the different structures of each ice.}
\label{fig:pt}
\end{figure}

The apparent differences in the ice phases observed in this work using the TIP4P/2005 model, and by Han {\em et al.}~\cite{han2010} and Bai and Zeng~\cite{bai2012} using the TIP5P model, could be attributed to the different force fields used. However, closer inspection reveals striking similarities between the disordered triangular phase described here, the high-density rhombic phase described by Han {\em et al.}~\cite{han2010}, and the high-density amorphous phase described by Bai and Zeng~\cite{bai2012}. The confusion stems from the fact that individual snapshots appear quasi-amorphous due to the fluctuating distortions on the lattice caused by the high-entropy proton disorder~\cite{corsetti2015b}; it is also difficult to establish the symmetry of the lattice for the same reason. Nevertheless, the snapshots and RDFs shown in these previous studies strongly suggest that the same phase is being observed in all these works. This, together with the good agreement found with density-functional theory~\cite{corsetti2015b}, lends support to the findings of this work beyond the particular force field used.

\subsection{Phase transitions}
The large structural and dynamical changes and the coexistence area (see Appendix~\ref{A1}) shown in Fig.~\ref{fig:pd} suggest that the phase transition from liquid to honeycomb ice at low densities is a first-order phase transition as previously reported~\cite{han2010}. We calculate the potential energy of the system as a function of density in order to investigate the nature of the phase transitions that occur at high densities. For this purpose, we take as the initial state the final configuration obtained in the previous MD simulations at $\rho = 1.37$ g~cm$^{-3}$ and $T = 240, 300$ K. We then increase the density by reducing the size of the cell along the {\em xy} plane in many steps of $\Delta$$\rho$ $= 0.01$ g~cm$^{-3}$ each. Between each $\Delta$$\rho$ step, we run 5 ns of re-equilibration with the Berendsen thermostat, followed by 100 ps of {\em NVE} statistics.The paths followed on the phase diagram are shown by the two arrows labeled 1a and 1b in Fig.~\ref{fig:pd}. The results are shown in Fig.~\ref{fig:pt}. At $T = 240$ K and $\rho$ $= 1.43$ g~cm$^{-3}$ we observe a change in potential energy of 8.3 meV per molecule that, together with the coexistence area in Fig.~\ref{fig:pd}, clearly indicate a first-order phase transition between the liquid and the square tubes ice. Instead, at $T = 300$ K, we do not observe any distinguishable energy jump related with a phase transition, although there is a clear change in the structure as the density is increased (see insets in Fig.~\ref{fig:pt} and other indicators in Appendix~\ref{A2}). This suggests that the liquid-hexatic and hexatic-triangular ice phase transitions are continuous, which would correspond to the continuous transition reported by Han {\em et al.}~\cite{han2010}.

\begin{figure}
\subfigure[\ ]{\includegraphics[width=0.49\textwidth]{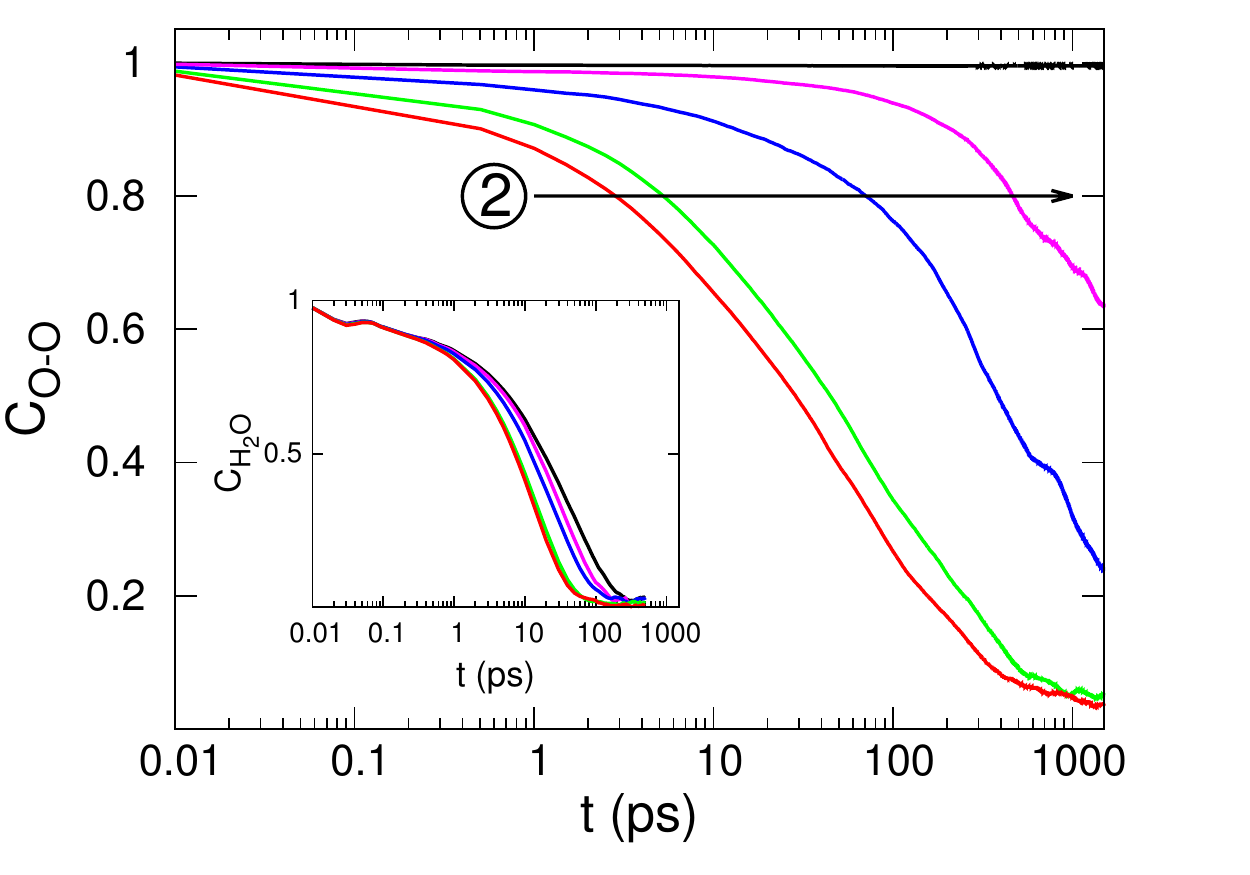}}
\subfigure[\ ]{\includegraphics[width=0.2\textwidth]{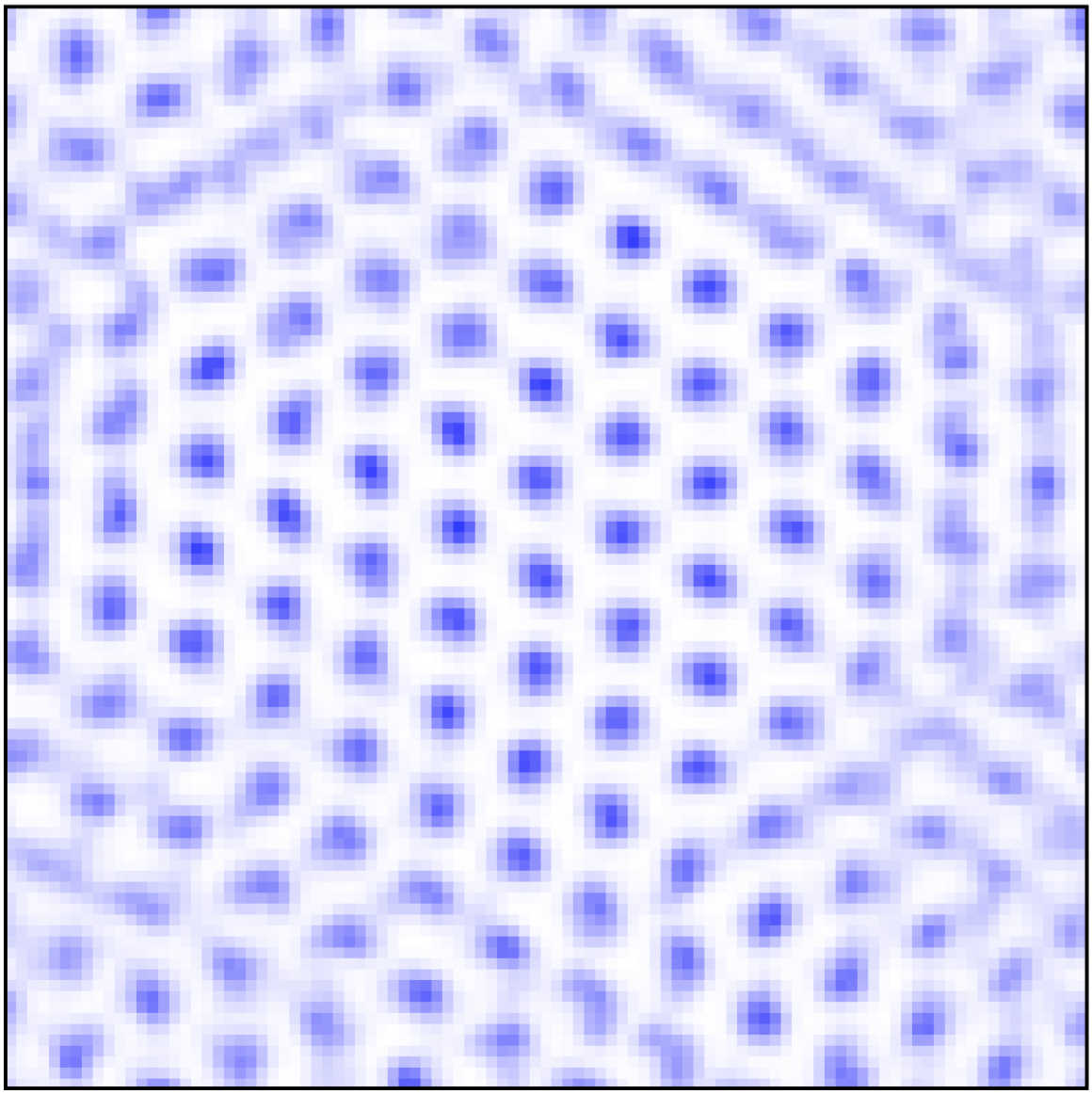}}\hspace{2.0 mm}
\subfigure[\ ]{\includegraphics[width=0.2\textwidth]{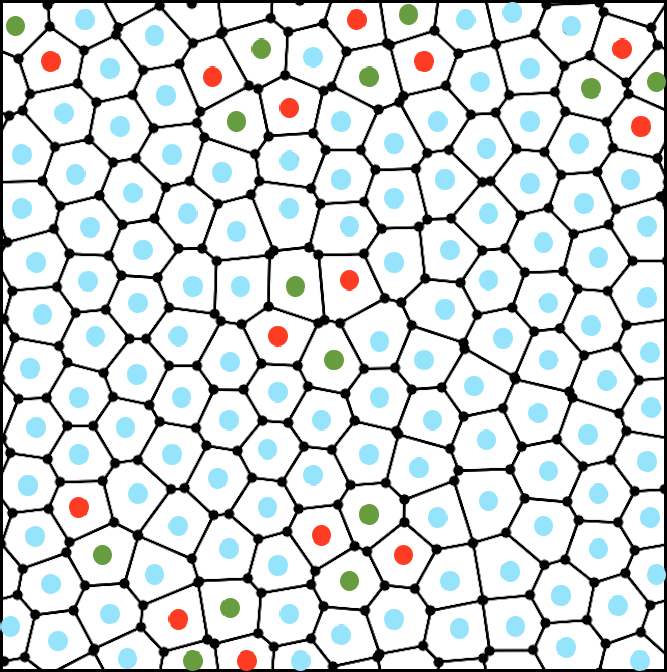}}
\caption{(a) Oxygen-oxygen first-neighbor correlation function (main plot) and dipole-dipole autocorrelation function (inset) as a function of time at  $T = 300$ K and five different densities: $\rho$ $= 1.27$ (red), $1.37$ (green), $1.47$ (blue), $1.54$ (pink), and $1.57$ (black) g~cm$^{-3}$. The path followed on the phase diagram is the one shown by the arrow labeled 2 in Fig.~\ref{fig:pd}. (b) Averaged positions of the oxygens during 100 ps and (c) Voronoi diagram of the oxygens located in the lower layer at a particular time for $\rho$ $= 1.47$ g~cm$^{-3}$ and $T$ $= 300$ K. Each type of polygon in the diagram is marked by different colors: red (pentagon), blue (hexagon), and green (heptagon).}
\label{fig:cor}
\end{figure}

One of the main characteristics of the triangular ice we observe is that the position of the oxygens are well fixed in closed-packed planes, while the hydrogens show a large disorder, which gives rise to a high configurational entropy~\cite{corsetti2015b}. Therefore, during the phase transition from liquid to triangular ice, we can expect a transition for oxygens but no noticeable change for hydrogens. To verify this, we calculate the oxygen-oxygen first-neighbor correlation function (C$_{\text{O-O}}$) and the dipole-dipole autocorrelation function (C$_{\text{H$_{2}$O}}$). For the former, we calculate the proportion of initial in-plane nearest-neighbor oxygens of any oxygen that remain after time {\em t}. Both correlation functions are averaged over all molecules and different initial times. Within the proposed scenario, after the freezing of the oxygens, C$_{\text{O-O}}$ would remain at a value close to 1, while C$_{\text{H$_{2}$O}}$ would decay to 0 due to the random motion of the hydrogens.

\begin{figure}[t]
\includegraphics[width=0.49\textwidth]{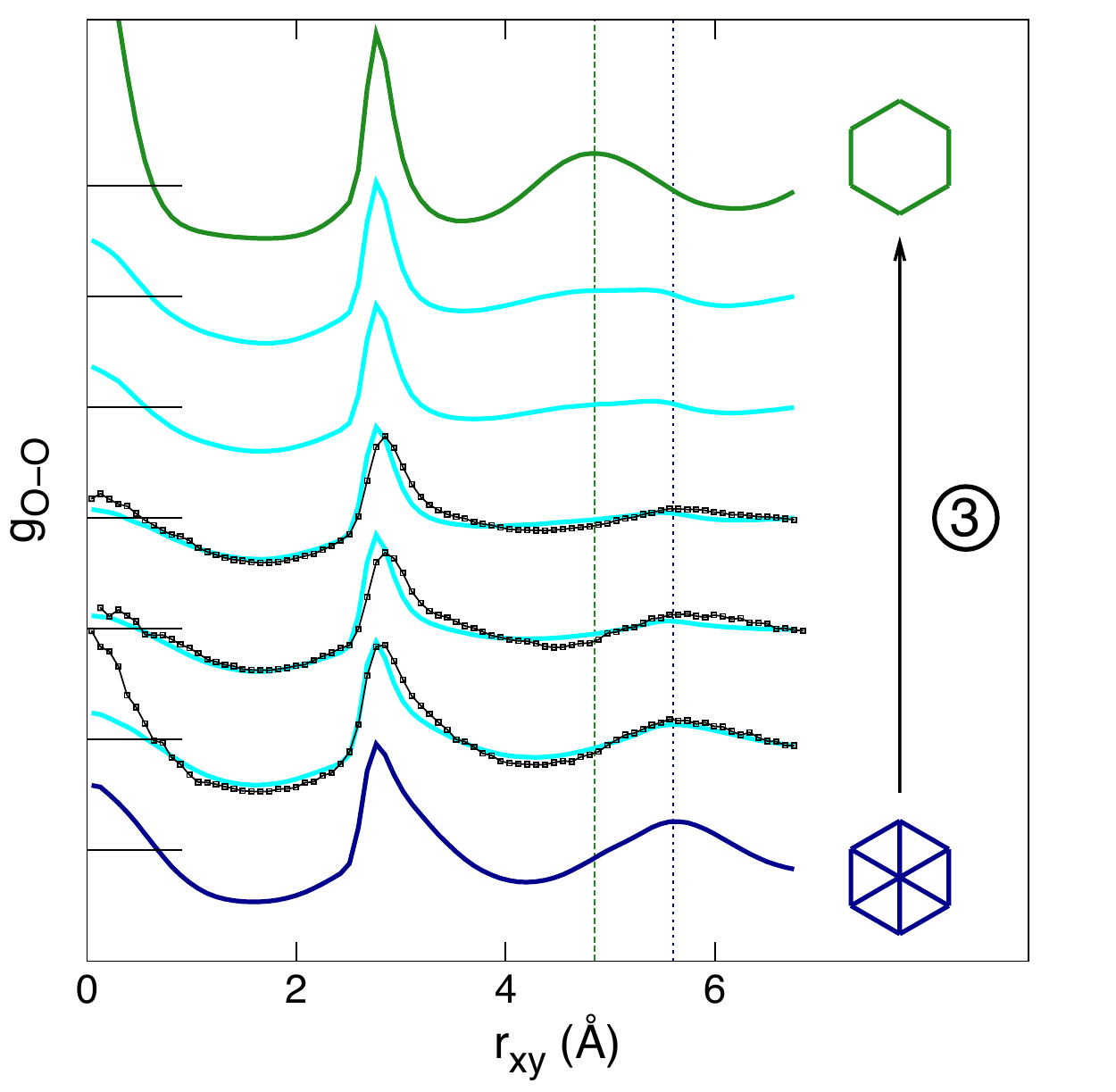}
\caption{Oxygen-oxygen radial distribution functions at different densities and temperatures following the path marked by the arrow labeled 3 in Fig.~\ref{fig:pd}. The curves are shifted on the {\em y} axis; the value of saturation of each curve is marked by a horizontal finite line. The smooth-line curves correspond to the MD calculations while the line-point curves to the AIMD calculations. The vertical dashed lines highlight the position of the second neighbor peak for the triangular ice (dark blue) and the honeycomb ice (green).}
\label{fig:rdf}
\end{figure}

Figure~\ref{fig:cor}(a) shows the correlation functions obtained at five different densities at $T = 300$ K (see Fig.~\ref{fig:pd}, arrow 2): Two of them are in the liquid phase ($\rho$ $= 1.27$, 1.37 g~cm$^{-3}$), another two are in the hexatic phase ($\rho$ $= 1.47$, 1.54 g~cm$^{-3}$), and one is in the triangular ice phase ($\rho$ $= 1.57$ g~cm$^{-3}$). The decay profile for C$_{\text{H$_{2}$O}}$ decays smoothly to zero for all five densities, showing no appreciable change when crossing the phase transitions; for C$_{\text{O-O}}$, however, there is a noticeable difference between the liquid (fast decay), the hexatic (slow decay), and the triangular ice phase (no decay). This result shows that only the oxygen atoms undergo the phase transitions, while the hydrogens remain in a quasi liquid dynamically disordered state. This behavior is analogous to the one observed in the plastic crystal phases obtained in bulk water at high pressures using the TIP4P/2005 model: The water molecules show large orientational disorder~\cite{aragones2009a,aragones2009b}.

To prove that the phase at $\rho$ $= 1.47$, 1.54 g~cm$^{-3}$, and $T = 300$ K is hexatic, we analyze the positions of the oxygens during the run: We observe that although there is a clear organization shown in the averaged positions of the oxygens, there is shear motion along the main directions of the triangular lattice [Fig.~\ref{fig:cor}(b)]. These anisotropic movements of the oxygens, as well as explaining the slow decay of C$_{\text{O-O}}$, suggest that the oxygen lattice has orientational long-range order but no translational long-range order, which is precisely what characterizes the hexatic phase. 

The KTHNY theory~\cite{kosterlitz1973,young1979,halperin1978,nelson1979} predicts continuous phase transitions for 2D materials in which an intermediate hexatic phase is located between the isotropic liquid and the crystalline solid phases. This theory is based on the creation and disassociation of dislocations~\cite{von2007}. In order to support our observation of the hexatic phase, we search for the presence of dislocations in the oxygen lattice. Figure~\ref{fig:cor}(c) shows the Voronoi diagram obtained from the position of the oxygens located in the lower layer at a given time. We observe that although most of the diagram is made of hexagons, there are pentagon-heptagon defect pairs that are characteristic of the presence of dislocations within a triangular lattice~\cite{von2007}. The Voronoi diagrams obtained at different times (see Appendix~\ref{A3}) show that these dislocations move along the triangular lattice during the run, explaining the fuzziness of some of the system due to shear shown in Fig.\ref{fig:cor}(b). Although there is another theory that describes the melting in two dimensional systems via the spontaneous generation of grain boundaries~\cite{chui1983}, in this case, all the results strongly suggest the existence of an intermediate hexatic phase at high densities and temperatures on the phase diagram (light purple area in Fig.~\ref{fig:pd}), and that the phase transition follows the KTHNY theory, as observed for a single layer (continuous phase transition) and double layer (weakly first-order phase transition) of Lennard-Jones particles~\cite{radhakrishnan2002}. The question of whether the observed phase transition is strictly continuous or very weakly first order cannot be answered with certainty from current results and is left for future studies.

Concerning the origin of the difference between the two melting processes (one continuous, one discontinuous), it could be inferred from the previous discussion that it is stemming from the H-disorder kept in the hexatic and triangular phases, as opposed to the square-shape tubes (and honeycomb) phase. Indeed, the thermal delocalization in the hexatic and triangular phases implies both, an effective monoatomic system, giving rise to a 2D close-packing in each layer, and an effective screening of the electrostatic interactions among water molecules.

\subsection{Characterization of liquid}
After analyzing the different phase transitions, we focus on the characterization of the liquid. Some of the properties of the liquid agree well with previously reported works~\cite{zangi2004}: The oxygens are organized into two main layers that are bridged by a constant flux of molecules. When the density of the liquid is increased, these two main layers become more pronounced and the flux of molecules is reduced. The diffusivity of the liquid in the {\em xy} plane is similar to the one of bulk water ($D \sim10^{-5}$ cm$^2$~s$^{-1}$ at $T = 300$ K). 

\begin{figure}[t]
\includegraphics[width=0.49\textwidth]{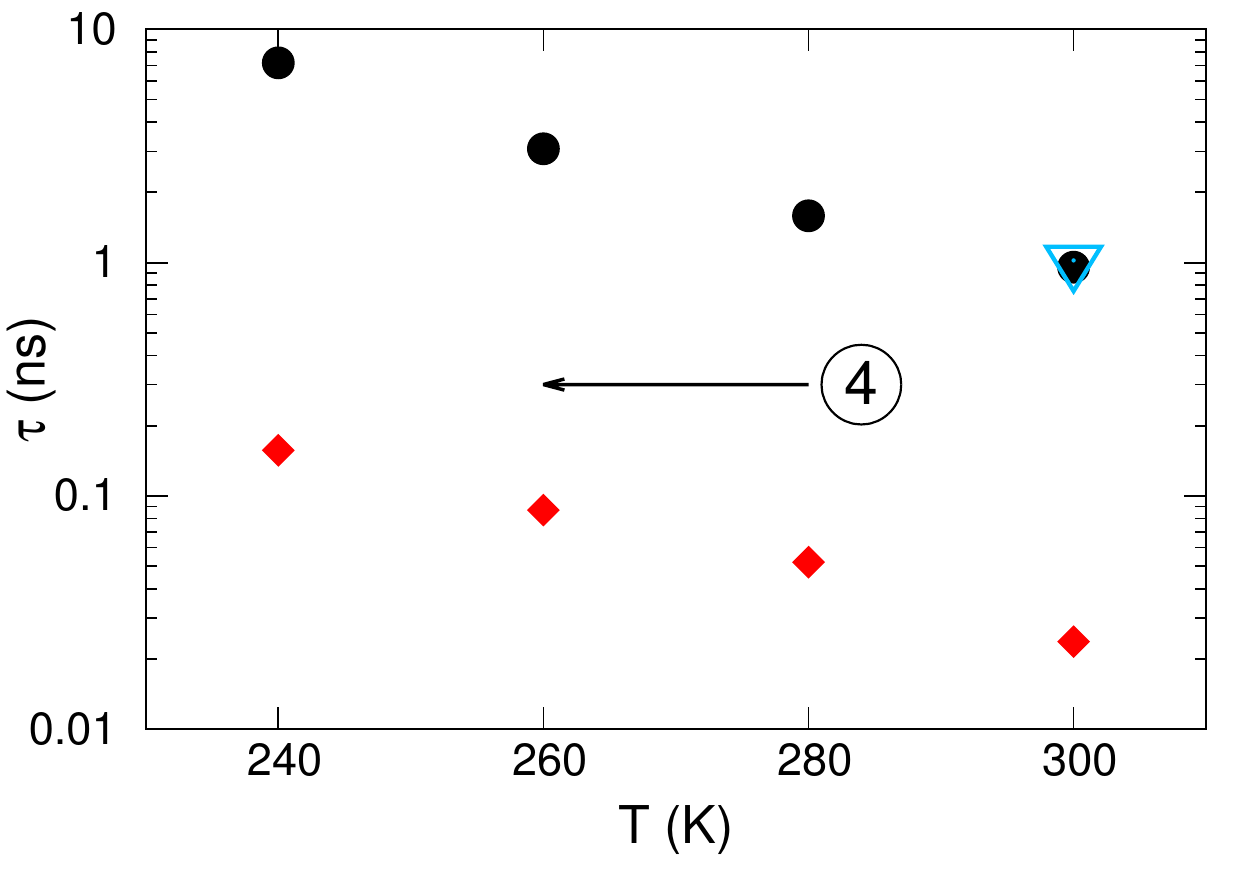}
\caption{Vertical-jump lifetime ($\tau_{\text{z}}$) obtained from MD (circles) and AIMD (triangle) calculations and the in plane-jump lifetime ($\tau_{\text{xy}}$, rhonbuses) at $\rho$ $= 1.37$ g~cm$^{-3}$ and different temperatures (arrow 4 in Fig.~\ref{fig:pd}).}
\label{fig:life}
\end{figure} 

In order to characterize the structure of the liquid, we take as reference the honeycomb and triangular ices (the square tubes ice is a particular state of the triangular ice~\cite{corsetti2015b}) and we check if the liquid exhibits the characteristics of either of these two solids. We choose a path within the phase diagram that connects the two phases (see arrow 3 in Fig.~\ref{fig:pd}) and we analyze the RDFs at each calculated point along this path. Figure~\ref{fig:rdf} shows the RDFs obtained from MD and AIMD calculations. Although the water obtained from AIMD tends to be more structured than the one obtained by MD, the two are in reasonable agreement for the purposes of this study. One clear difference between the RDF of the honeycomb and triangular phases is the position of the second neighbor peak: For the honeycomb phase it is at $r_{\text{h}} = \sqrt{3}a$, where {\em a} is the first neighbor distance, while for the triangular phase it is at $r_{\text{t}} = 2a$. It is important to note that the peak at $r_{\text{xy}} = \sqrt{3}a$ is also present for the triangular phase; however, the large oxygen lattice vibrations caused by the hydrogen disorder broaden it out sufficiently for it to be no longer distinguishable. Although we observe a continuous shifting of the second neighbor peak from $r_{\text{t}}$ towards $r_{\text{h}}$ as we get closer to the honeycomb ice RDF in Fig.~\ref{fig:rdf}, it is very significant how all the RDFs coming from the liquid samples show the characteristic peak of the triangular phase at $r_{\text{t}}$, suggesting that the liquid maintains the local structure of triangular phase. 

The first maximum at $r_{\text{xy}} = 0$ measures the correlation between the molecules at the same {\em xy} position but in different layers. As all the RDFs in Fig.~\ref{fig:rdf} show a pronounced peak at this position, we conclude that there is a strong correlation between the two layers, with a strong tendency for every O in one layer to have another one just across in the other layer. This correlation is increased by increasing the density or decreasing the temperature.

\begin{figure}[t]
\includegraphics[width=0.49\textwidth]{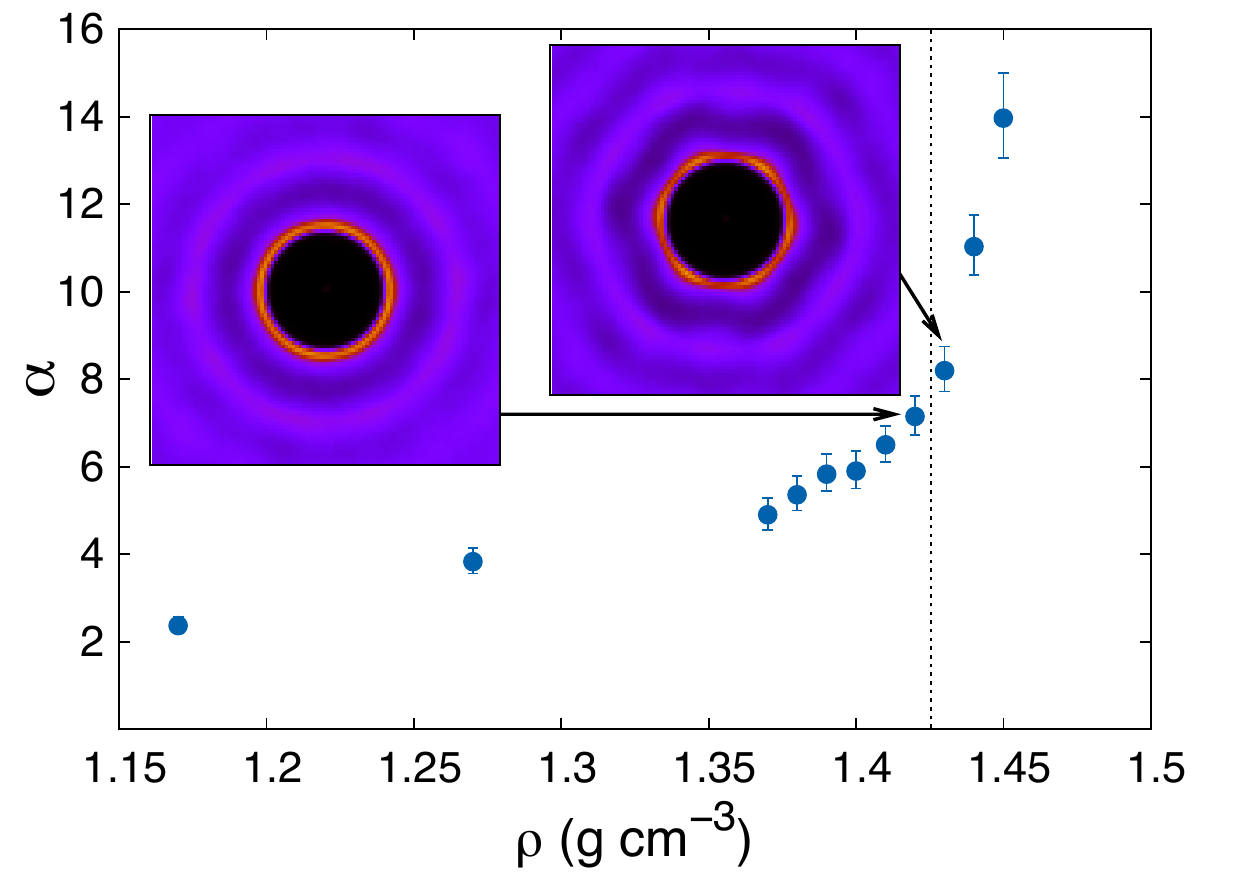}
\caption{Ratio between the characteristic times of C$_{\text{O-O}}$ and C$_{\text{H$_{2}$O}}$ at $T = 300$ K with respect to density (main plot) and 2D oxygen-oxygen correlation function (insets) at $\rho$ $= 1.42$ (left) and $1.43$ g~cm$^{-3}$ (right): 
the brighter the color, the larger the value of the 2D oxygen-oxygen correlation function. The vertical dashed line represents the estimated density in which the liquid-hexatic transition occurs.}
\label{fig:factor}
\end{figure} 

Our various indicators suggest that the high-density liquid maintains many of the characteristics of triangular ice: high interlayer correlation and a local triangular structure of the oxygens, as shown in the RDFs, and a distinct separation between the two layers, as shown in the density profiles (see Appendix~\ref{A4}). In order to verify the latter, we estimate the average time ($\tau_{\text{z}}$) that a molecule stays within a layer before jumping to the other layer. To do so, we divide the cell along the confining direction into two equal parts, and we extract the number of jumps that have occurred during the run from one side of the cell to the other. Second, we estimate from the calculated diffusivity $D$, the time needed by a molecule to jump a distance {\em a} within one layer (which coincides with the interlayer distance) using a random walk model: $\tau_{\text{xy}} = a^{2}$(4$D$)$^{-1}$. Figure~\ref{fig:life} shows these two values at $\rho$ $= 1.37$ g~cm$^{-3}$ and different temperatures (following arrow 4 in Fig.~\ref{fig:pd}). The values of $\tau_{\text{z}}$ obtained from empirical MD and AIMD calculations at T $= 300$ K are almost identical, supporting the reliability of our calculations. In all cases, $\tau_{\text{z}}$ is around 50 times larger than $\tau_{\text{xy}}$, which shows that the velocity scale of the diffusion in {\em xy} is much larger than in {\em z}. These results, together with the high interlayer correlation shown by the RDF, confirm that a molecule remains for an average of 10 ns in one layer before jumping to the other layer, and that its in-layer motion is closely mirrored by a partner molecule in the other layer; in other words, the AA stacking is maintained by the liquid.

Furthermore, the decoupling in the dynamics between O and H that occurs in the hexatic and triangular phases is also clearly observed in the high-density liquid. Figure~\ref{fig:factor} shows the ratio ($\alpha$) with respect to density at $T =$ 300 K. This ratio is given by $\alpha$ $=$ C$_{\text{O-O}}(\tau_{\text{O-O}})$~C$_{\text{H$_{2}$O}}(\tau_{\text{H$_{2}$O}})^{-1}$, where the characteristic time $\tau$ of each correlation function is obtained from C($\tau$) $=$ 0.5. The larger the value of $\alpha$, the larger the decoupling between the dynamics of O and H. We estimate the density at which the liquid-hexatic phase transition occurs by looking at the 2D O-O correlation function (insets in Fig.~\ref{fig:factor}), which are analogous to the RDFs but take into account both the $x$ and $y$ coordinates of the oxygens instead of the radial coordinate. The 2D O-O correlation function at $\rho$ $= 1.42$ g~cm$^{-3}$ shows a spherically symmetric first-neighbor ring, characteristic of the liquid, while at $\rho$ $= 1.43$ g~cm$^{-3}$ it transforms into a hexagon, characteristic of the hexatic phase. Therefore, the phase transition at $T = 300$ K is estimated to occur at  $\rho$ $= 1.425$ g~cm$^{-3}$. The important point to note is the behavior of alpha: It increases not only within the hexatic phase, but also from low- to high-density liquid. This means that the decoupling between the dynamics of O and H already occurs within the liquid. This points to a regime constituted by the triangular, hexatic, and liquid phases, in which the complexity of bilayer water seems to disappear, resulting in what resembles a simple monoatomic fluid.

\section{Conclusions}

Our simulations for bilayer water indicate the presence of two types of melting at high densities: a first order phase transition into an ice made of square tubes, at low temperature, and a continuous phase transition into triangular ice at higher $T$. For the latter the observed phenomenology strongly suggests KTHNY two dimensional melting, including the observation of an intermediate hexatic phase between the solid and the liquid. During this continuous melting only the oxygens are affected, while the hydrogens keep behaving liquid-like, resulting in a unusual decoupling in the dynamics of each species.
 
The characterization of the liquid shows that the triangular local structure is maintained, and the two layers are strongly correlated with very infrequent exchange of matter. We observe that the decoupling between the dynamics of O and H already starts in the liquid phase, showing the existence of a regime in the phase diagram constituted by the triangular, hexatic, and liquid phases, in which water resembles a simple monoatomic fluid.

The unusual characteristics of the system in this regime allows us to expect that the dielectric properties of bilayer water differ markedly from the ones observed for its bulk counterpart. A preliminary estimation of the relative dielectric constant $\epsilon_r$ along the planar direction of the triangular ice shows that its value is slightly larger than 200, substantially higher than the 53 obtained at room pressure and $T = 273$ K for Ih ice using the TIP4P/2005 model~\cite{aragones2010}. Due to the large constraints along the confining direction, the out-of-plane component of $\epsilon_r$ is expected to be very low: for the liquid at $\rho$ $= 1.17$ g~cm$^{-3}$ and $T = 300$ K, we have obtained a value of $\epsilon_r$ = 3.1. Moreover, we estimate the Debye relaxation time of triangular ice to be clearly below the nanosecond scale, close to the relaxation time observed for bulk water (17 ps) and far from the one observed for Ih ice (2.2 \textmugreek s) at room conditions~\cite{artemov2014}. A more detailed dielectric characterization of bilayer water and ice seems to be a promising topic for further work.

\section{Acknowledgements}
This work was partly funded by Grant No. FIS2012-37549-C05 from the Spanish Ministry of Economy and competitiveness and Grant No. Exp.\ 97/14 (Wet Nanoscopy) from the Programa Red Guipuzcoana de Ciencia, Tecnolog\'{i}a e Innovaci\'{o}n, Diputaci\'{o}n Foral de Gipuzkoa. We thank Jos\'{e} M. Soler and Pablo Aguado for useful discussions. The calculations were performed on the Arina HPC cluster (Universidad del Pa\'{i}s Vasco/Euskal Herriko Unibertsitatea, Spain) and MareNostrum (Barcelona Supercomputing Center). SGIker (UPV/EHU, MICINN, GV/EJ, ERDF, and ESF) support is gratefully acknowledged.

\begin{appendices}
\section{First order phase transition: structural and dynamical analysis}
\label{A1}

\begin{figure}[h]
    \begin{center}
      \subfigure[\ ]{ \includegraphics[width=0.3\columnwidth]{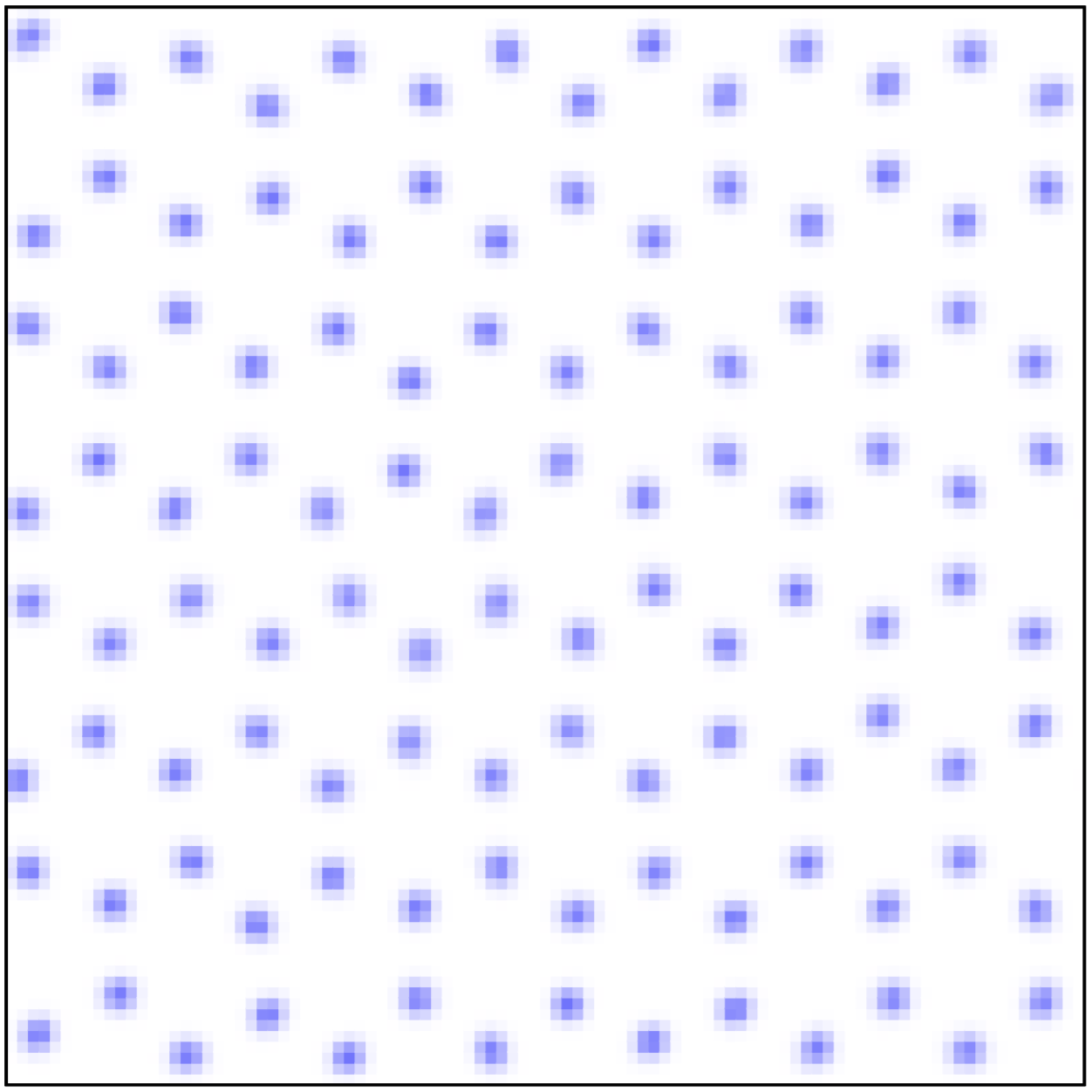}}
      \subfigure[\ ]{ \includegraphics[width=0.3\columnwidth]{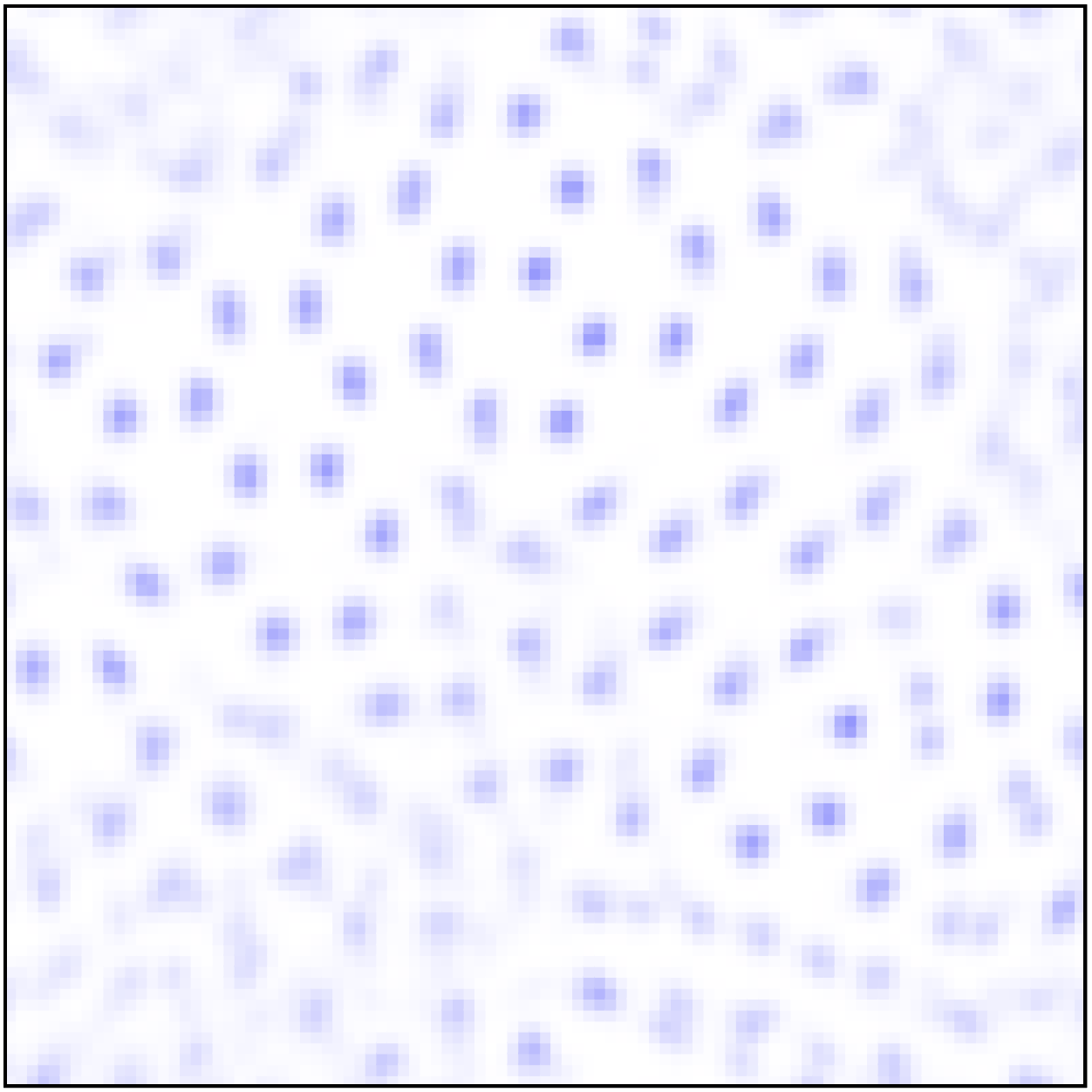}}
      \subfigure[\ ]{ \includegraphics[width=0.3\columnwidth]{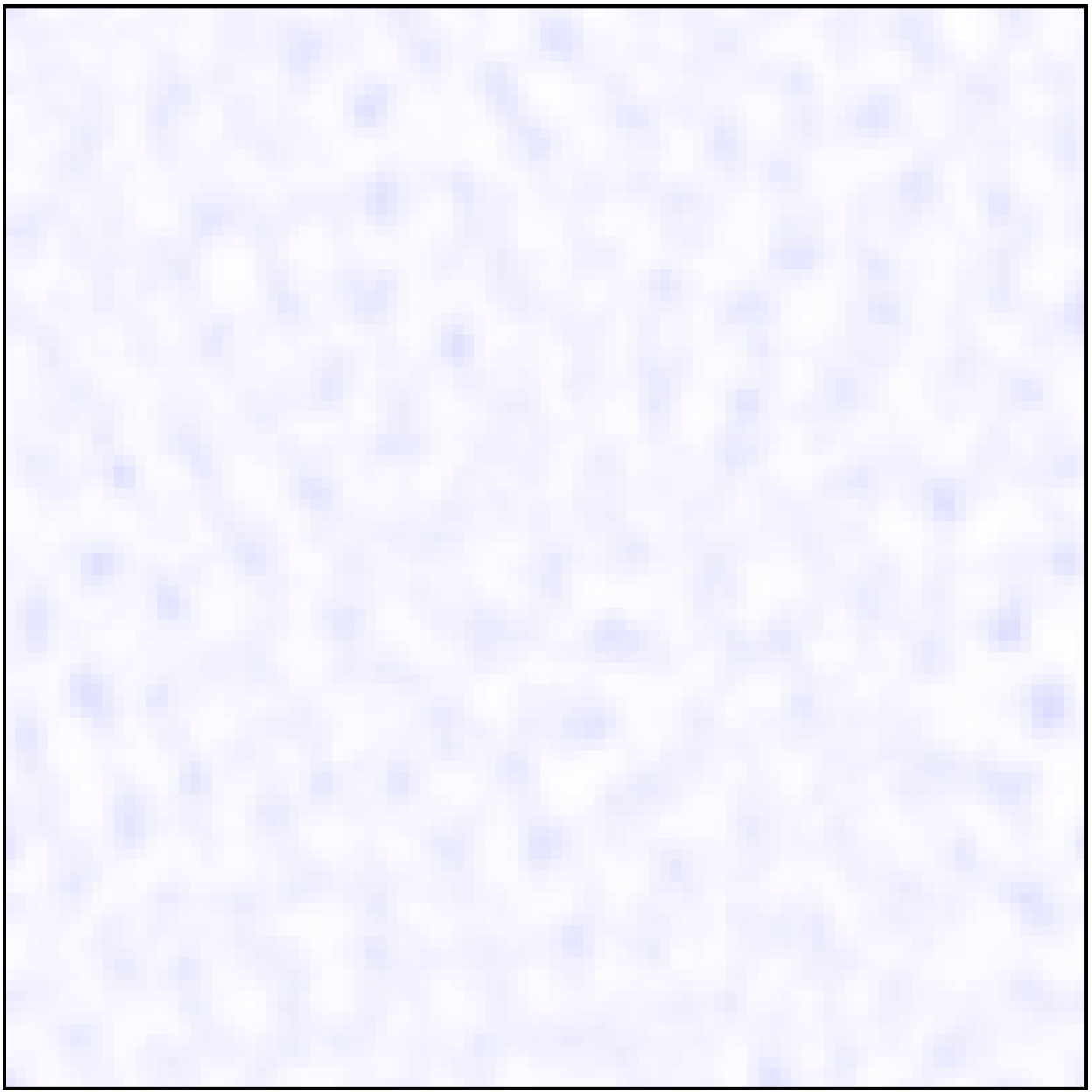}}\\
      \subfigure[\ ]{ \includegraphics[width=0.48\columnwidth]{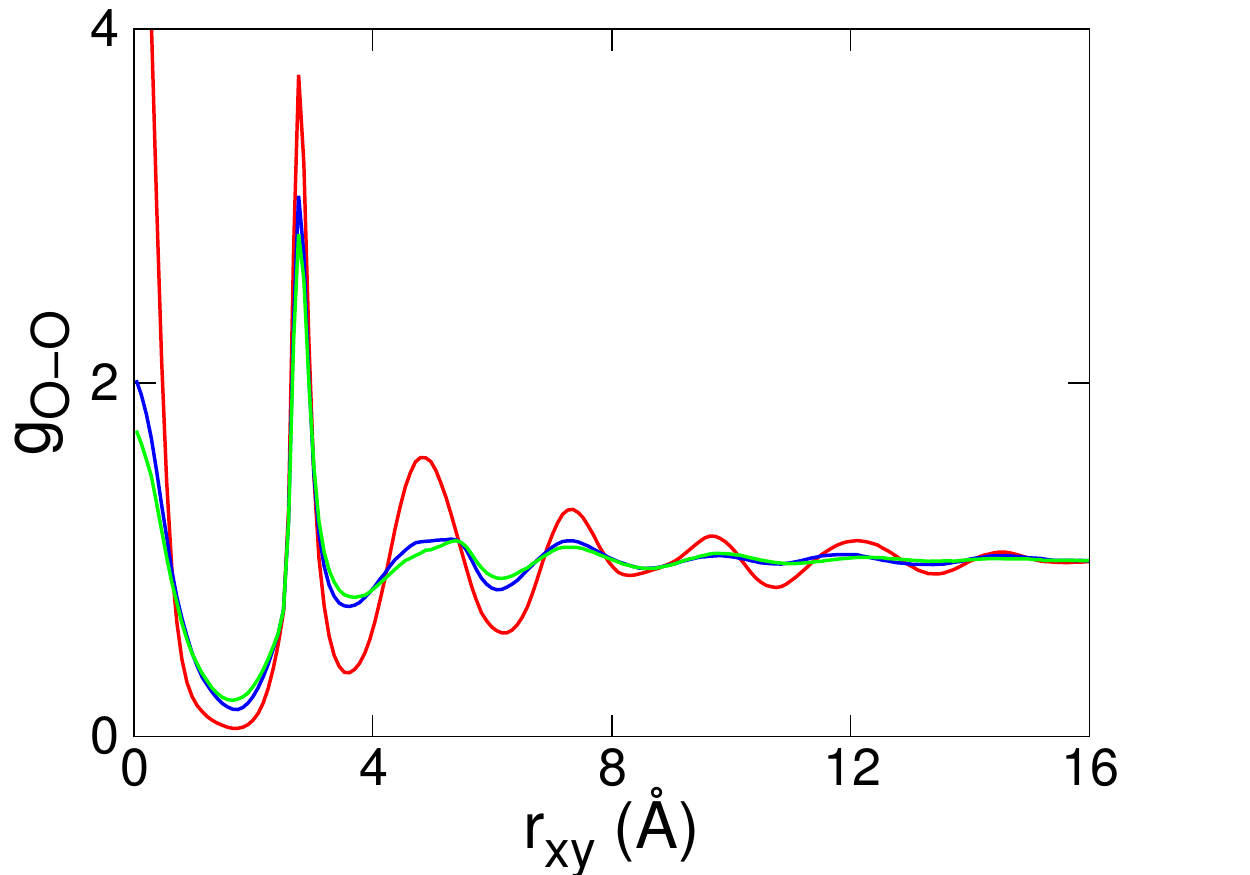}}
      \subfigure[\ ]{ \includegraphics[width=0.48\columnwidth]{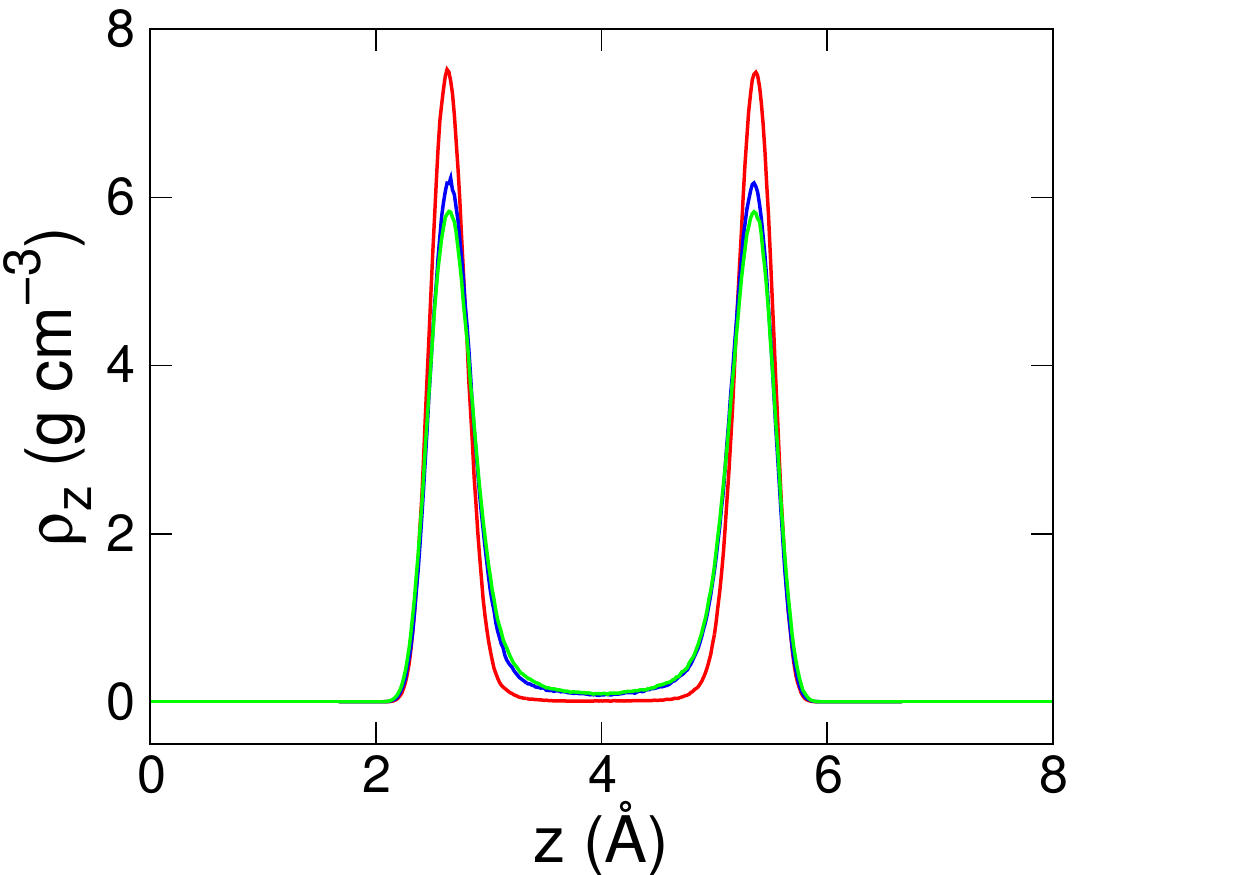}}\\
      \subfigure[\ ]{ \includegraphics[width=0.48\columnwidth]{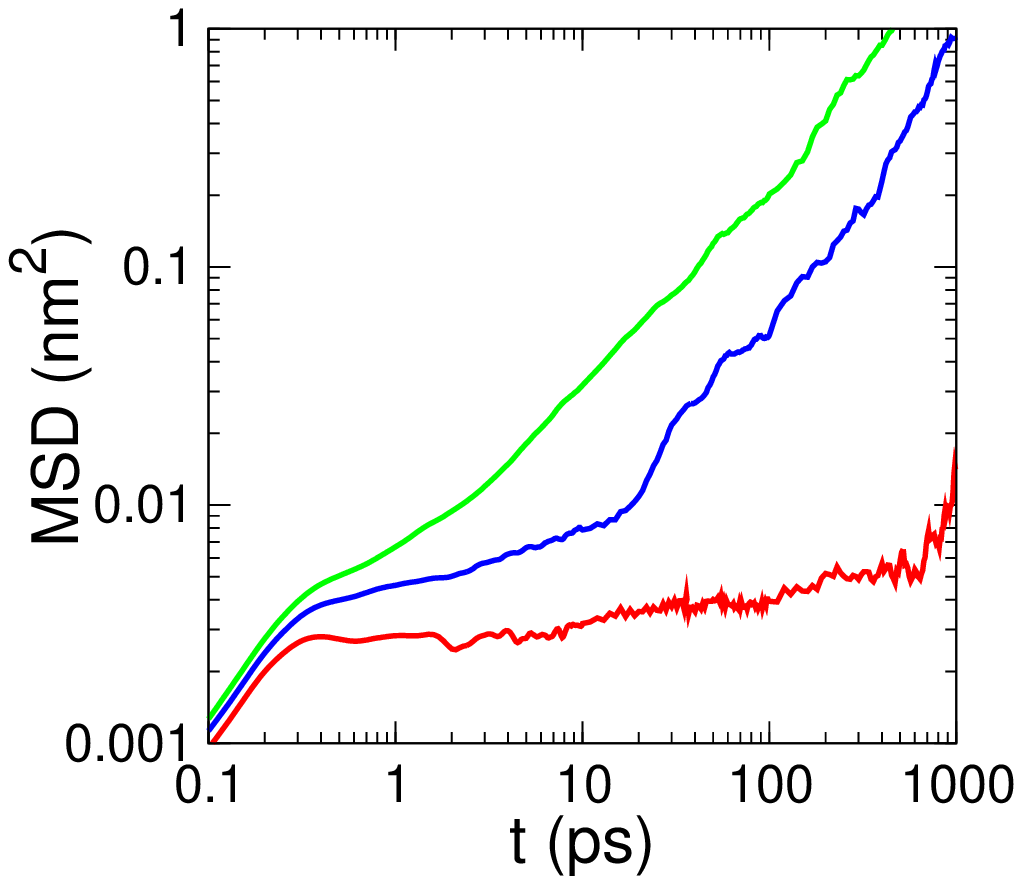}}

    \end{center}
    \caption{ [(a)-(c)] Averaged positions of the oxygens during 100 ps at $\rho$ $= 1.17$ g~cm$^{-3}$ and three
     different temperatures: (a) 240 K, (b) 260 K, and (c) 280 K. (d) In-plane oxygen-oxygen radial distribution 
     function, (e) density profile of the oxygens along the confining direction, and (f) in-plane mean 
     square displacement of the oxygens at $\rho$ $= 1.17$ g~cm$^{-3}$ and three
     different temperatures: 240 K (red), 260 K (blue), and 280 K (green).
           }
    \label{fig:s1}
 \end{figure}

The different indicators used in this work show areas with solid-liquid phase coexistence and large structural and dynamical changes during the phase transitions between the liquid and the two solids at low temperature (honeycomb ice and square-shape tubes ice) in the phase diagram. These results clearly indicate that there is a first order phase transition connecting the liquid with these two solids. Figure~\ref{fig:s1}
shows these indicators at three points in the phase diagram that connect the liquid with the honeycomb ice phase. The averaged positions of the oxygens at 260 K shows coexistence of both phases, ice and liquid [Fig.~\ref{fig:s1}(b)], located between the honeycomb ice [Fig.~\ref{fig:s1}(a)] and the liquid [Fig.~\ref{fig:s1}(c)]. The large structural and dynamical changes occurring during the phase transition can be observed in the oxygen-oxygen RDFs [Fig.~\ref{fig:s1}(d)], the density profiles of the oxygens [Fig~\ref{fig:s1}(e)] and the oxygen mean square displacement [Fig~\ref{fig:s1}(f)].

 \section{Continuous phase transition: structural and dynamical analysis}
\label{A2}

\begin{figure}[h]
    \begin{center}
      \subfigure[\ ]{ \includegraphics[width=0.45\columnwidth]{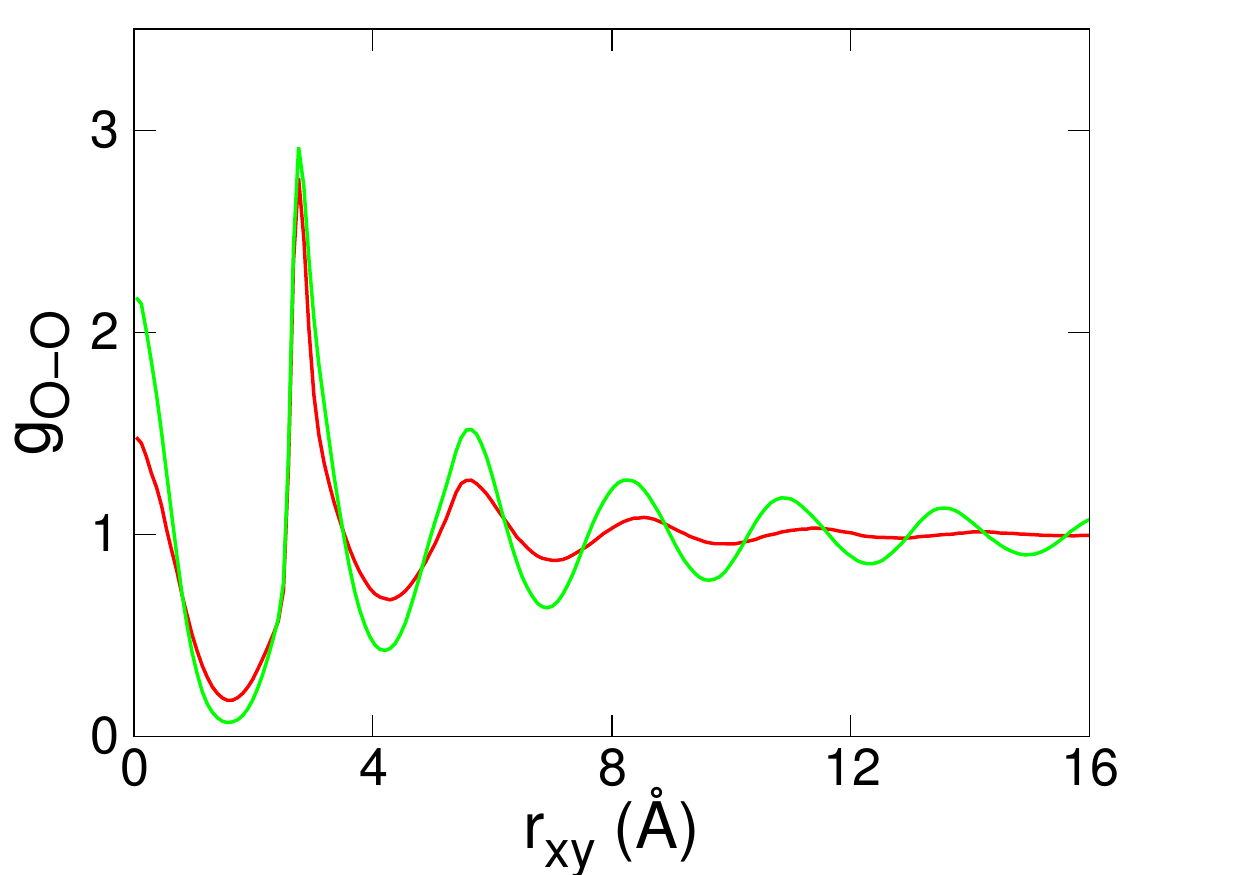}}
      \subfigure[\ ]{ \includegraphics[width=0.45\columnwidth]{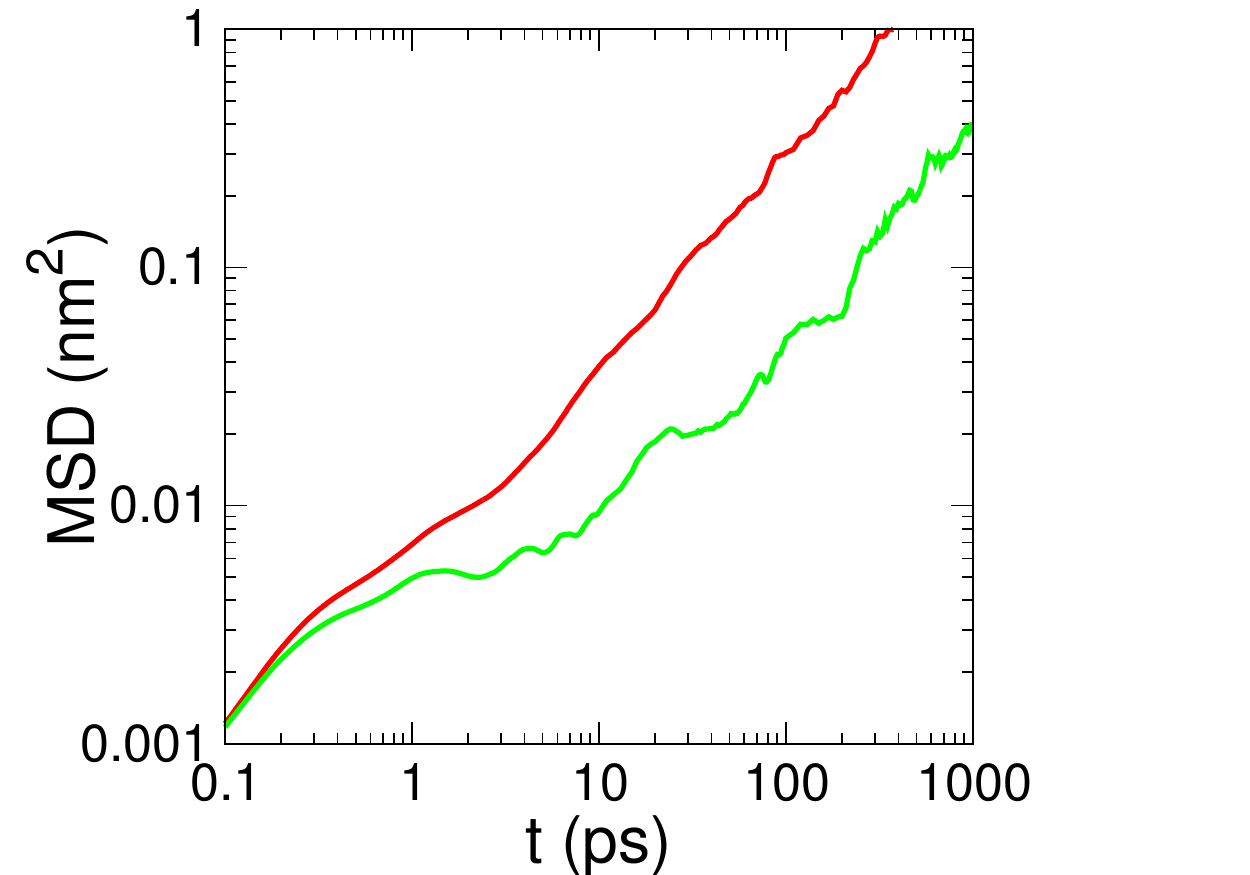}}
    \end{center}
    \caption{(a) In-plane oxygen-oxygen radial distribution function, and
    		(b) mean square displacement of the oxygens at $T$ $= 300$ K and 
		 $\rho$ $= 1.37$ (red), and $1.47$ (green) g~cm$^{-3}$}
    \label{fig:s2}
 \end{figure}

 The oxygen-oxygen RDFs obtained at $T$ $= 300$ K and 
 $\rho$ $= 1.37$ and $1.47$ g~cm$^{-3}$ [Fig.~\ref{fig:s2}(a)] agree
  with the results shown in the main text: Although we do not observe any clear change
  in the potential energy of the system while increasing density at 300 K (Fig.~2), the averaged
  positions of the oxygens (insets in Fig.~2) and the oxygen-oxygen RDFs clearly show
  a large change in the structure.
  
 The oxygen mean square displacements [Fig.~\ref{fig:s2}(b)] also show a change in the dynamical 
 behavior of water during the phase transition: for $\rho$ $= 1.37$ g~cm$^{-3}$, we obtain a diffusivity of $8.26~10^{-6}$ cm$^{2}~$s$^{-1}$, similar to the diffusivity of confined bilayer water at 300 K in similar conditions~\cite{zangi2004,han2010,kumar2005}. However, 
 for $\rho$ $= 1.47$ g~cm$^{-3}$, we obtained a diffusivity of $9.5~10^{-7}$ cm$^{2}~$s$^{-1}$, which
 is an intermediate value between the usual diffusivities that confined bilayer water ($\sim 10^{-5}$ cm$^{2}~$s$^{-1}$) and bilayer ice ($\sim 10^{-9}$ cm$^{2}~$s$^{-1}$) show at 300 K in similar conditions~\cite{zangi2004,han2010,kumar2005}. These results, together with the ones shown in the main text, support the existence of the intermediate hexatic phase at high temperatures.

\section{Voronoi diagrams at different times}
\label{A3}

\begin{figure}[h]
    \begin{center}
      \subfigure[\ ]{ \includegraphics[width=0.3\columnwidth]{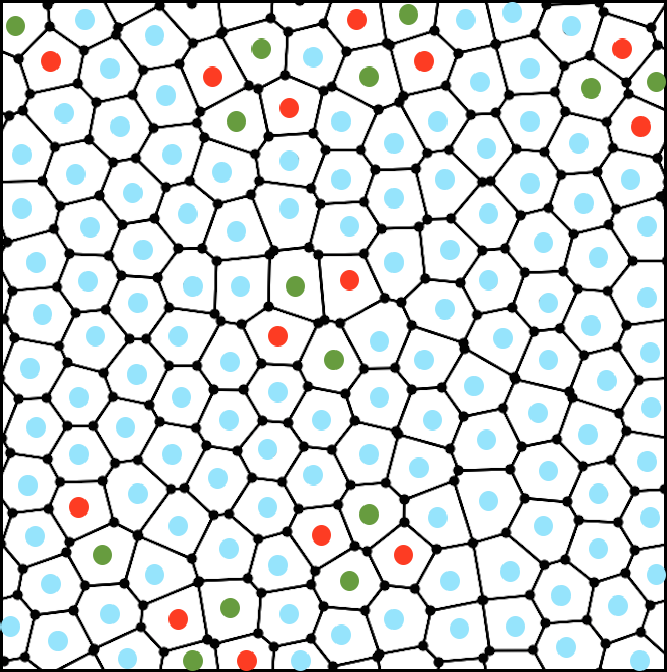}}
      \subfigure[\ ]{ \includegraphics[width=0.3\columnwidth]{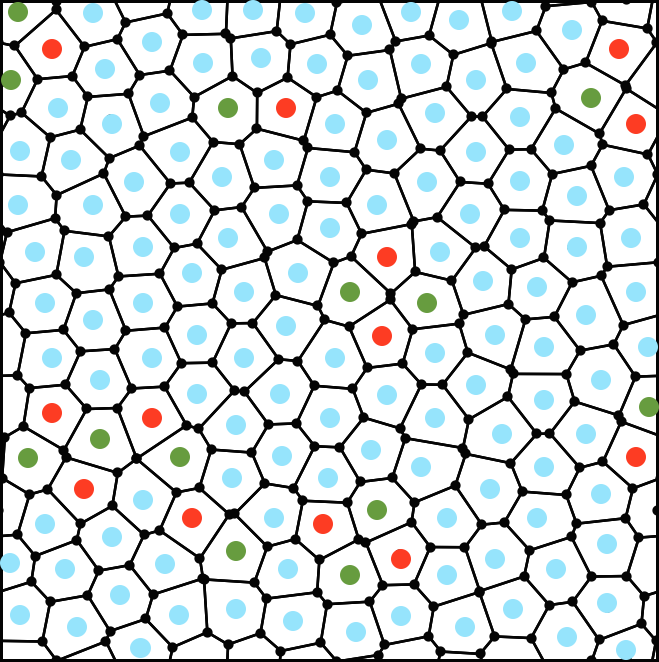}}
      \subfigure[\ ]{ \includegraphics[width=0.3\columnwidth]{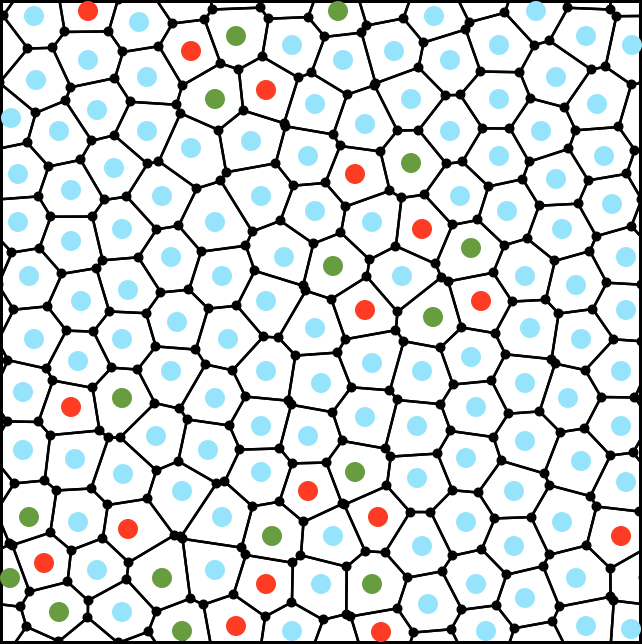}}
    \end{center}
    \caption{ [(a)-(c)] Voronoi diagram of the oxygens located  in the lower layer at
    $\rho$ $= 1.47$ g~cm$^{-3}$, $T$ $= 300$ K, and three different times: (a)
     0 ps, (b) 1 ps, and (c) 2 ps. Each type of polygon is marked by different colors:
     red (pentagon), blue (hexagon), and green (heptagon).
           }
    \label{fig:s3}
 \end{figure}

In order to study the diffusion of the dislocations within the hexatic phase, we plot the Voronoi diagram of the oxygens located in the lower layer at three different times (Fig.~\ref{fig:s3}). We observe that the dislocations (pentagon-heptagon pair defects) change their position between these times, explaining the shear moves shown in Fig.~3(b) and the slow decay of C$_{\text{O-O}}$ [Fig.~3(a)] of the hexatic phase. In all the three diagrams we observe the existence of an isolated pentagon-heptagon pair (single dislocation) at different postions. This particular defect is known to be responsible for breaking the translational long-range order within a triangular lattice~\cite{von2007}.
 
 \section{high-density liquid structured along the confining direction}
 \label{A4}
 
    \begin{figure}[t]
    \begin{center}
      \subfigure[\ ]{ \includegraphics[width=0.48\columnwidth]{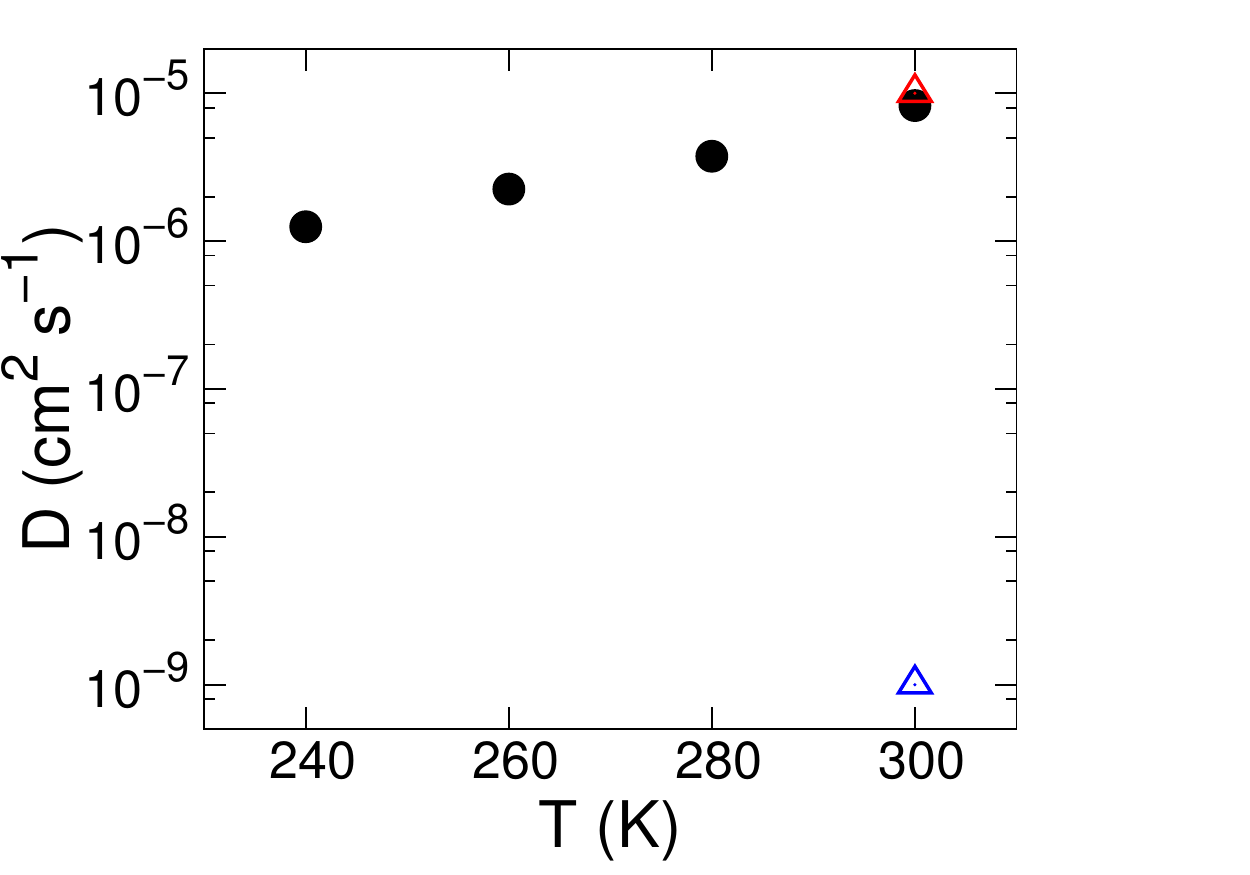}}
      \subfigure[\ ]{ \includegraphics[width=0.47\columnwidth]{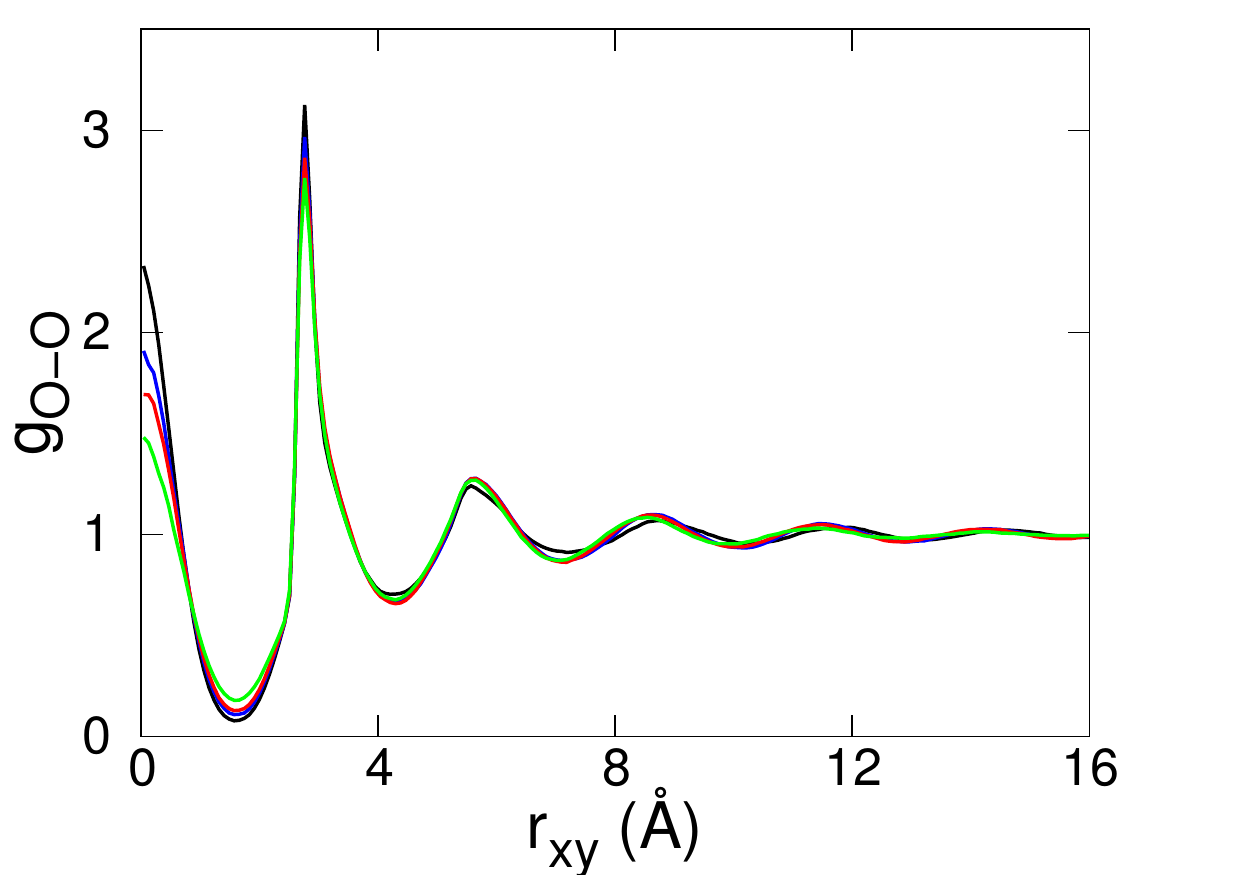}}\\
      \subfigure[\ ]{ \includegraphics[width=0.48\columnwidth]{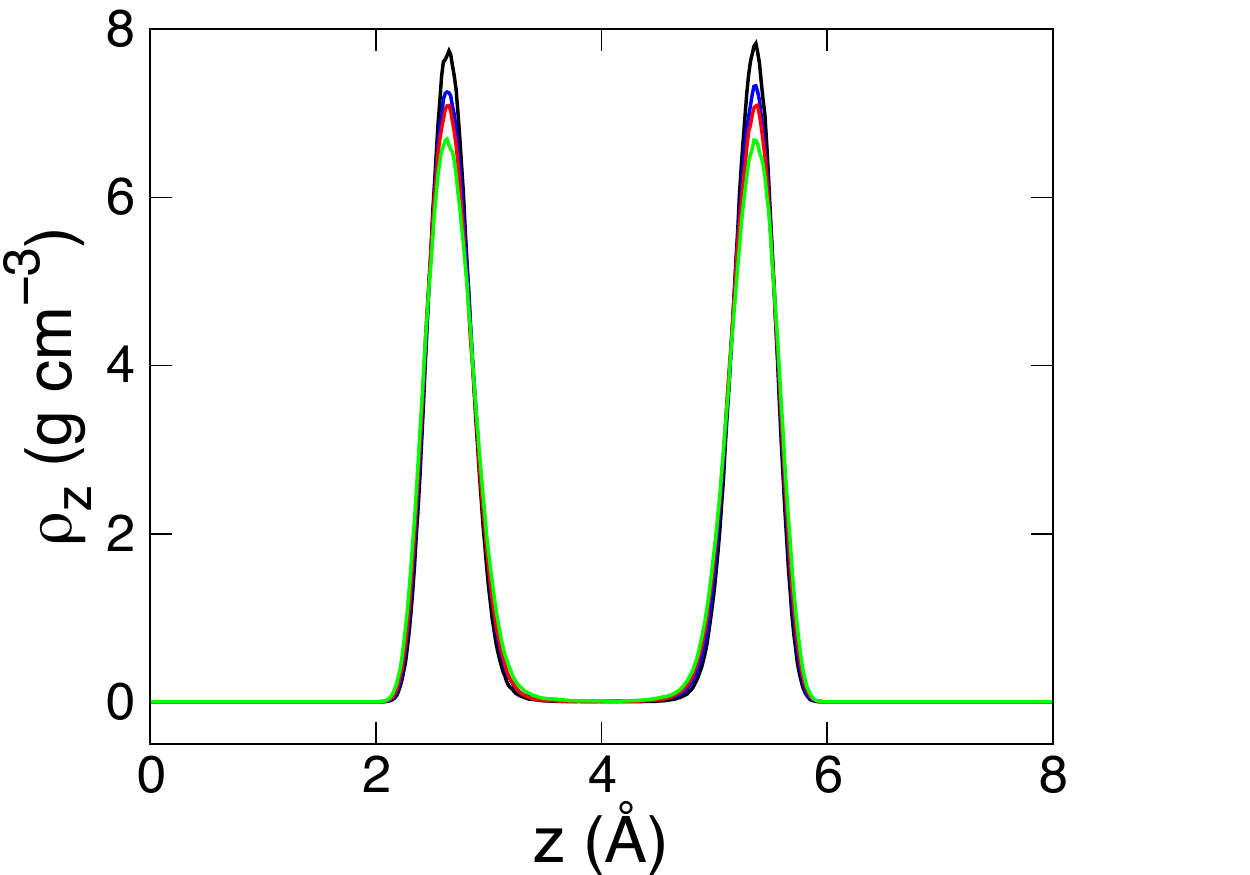}}
    \end{center}
    \caption{(a) In-plane diffusion constant of the oxygens at $\rho$ $= 1.37$ g~cm$^{-3}$. The expected 			value of the diffusion constant for a confined bilayer water (red triangle) and bilayer ice (blue 			triangle) are shown for comparison.
    		(b) In-plane oxygen-oxygen radial distribution function, and (c) density profile of the oxygens 
		along the confining direction, both at $\rho$ $= 1.37$ g~cm$^{-3}$ and four different 
		temperatures: 240 K (black), 260 K (blue), 280 K (red), and 300 K (green).}
    \label{fig:s4}
 \end{figure}

 The structural and dynamical analysis of the liquid at $\rho$ $= 1.37$ g~cm$^{-3}$ shows that in
 the {\em xy} plane, water behaves like a normal liquid, while along the confining direction it is highly structured. The in-plane oxygen diffusion constants shown in Fig.~\ref{fig:s4}(a) are very similar to the expected value of the diffusion constant for a confined bilayer water ($\sim 10^{-5}$ cm$^{2}~$s$^{-1}$) 
 at 300 K in similar conditions~\cite{zangi2004,han2010,kumar2005} and the in-plane oxygen-oxygen RDFs show a very similar liquid like structure for the four
 different temperatures (Fig.~\ref{fig:s4}(b)). The oxygen density profiles along the confining direction however, show that the molecules are structured in two main layers with almost no flux between them 
 [Fig.~\ref{fig:s4}(c)]. The pronounced peak at the origin of the RDFs show a high 
 correlation between molecules from different layers. 
  
\end{appendices}

\section{References}

\end{document}